\documentclass{emulateapj}
\usepackage{graphicx,amsmath,amssymb}
\usepackage{natbib}
\usepackage{apjfonts}

\bibpunct[, ]{(}{)}{;}{a}{}{,}
\begin{document}

   \title{Outshining the quasars at reionisation: \\The X-ray spectrum and lightcurve of the redshift~6.29 $\gamma$-Ray Burst GRB\,050904}

   \author{D.~Watson,\altaffilmark{1}
           J.~N.~Reeves,\altaffilmark{2,3}
           J.~Hjorth,\altaffilmark{1}
	   J.~P.~U.~Fynbo,\altaffilmark{1}
           P.~Jakobsson,\altaffilmark{1}
           K.~Pedersen,\altaffilmark{1}
           J.~Sollerman,\altaffilmark{1}
           J.~M.~Castro~Cer\'on,\altaffilmark{1}
           S.~McBreen,\altaffilmark{4}
       and S.~Foley\altaffilmark{5}
            }
   \altaffiltext{1}{Dark Cosmology Centre, Niels Bohr Institute, University of Copenhagen, Juliane Maries Vej 30, DK-2100 Copenhagen \O, Denmark; darach, jens, jfynbo, pallja, kp, jesper @astro.ku.dk, josemari@alumni.nd.edu}
   \altaffiltext{2}{Laboratory for High Energy Astrophysics, Code 662, NASA Goddard Space Flight Center, Greenbelt, MD 20771, USA; jnr@milkyway.gsfc.nasa.gov}
   \altaffiltext{3}{Dept. of Physics and Astronomy, Johns Hopkins University, 3400 North Charles St., Baltimore, MD 21218, USA}
   \altaffiltext{4}{Astrophysics Missions Division, Research Scientific Support Department of ESA, ESTEC, Noordwijk, The Netherlands; smcbreen@rssd.esa.int}
   \altaffiltext{5}{Dept.\ of Physics, University College Dublin, Dublin 4, Ireland; sfoley@bermuda.ucd.ie}
   \begin{abstract}

Gamma-ray burst (GRB) 050904 is the most distant X-ray source known, at
$z=6.295$, comparable to the farthest AGN and galaxies. Its X-ray flux
decays, but not as a power-law; it is dominated by large variability from a
few minutes to at least half a day. The spectra soften from a power-law with
photon index $\Gamma=1.2$ to 1.9, and are well-fit by an absorbed power-law
with possible evidence of large intrinsic absorption. There is no evidence
for discrete features, in spite of the high signal-to-noise ratio. In the
days after the burst, GRB\,050904 was by far the brightest known X-ray
source at $z>4$.  In the first minutes after the burst, the flux was
$>10^{-9}$\,erg\,cm$^{-2}$\,s$^{-1}$ in the 0.2--10\,keV band, corresponding
to an apparent luminosity $>10^5$ times larger than the brightest AGN at
these distances.  More photons were acquired in a few minutes with
\emph{Swift}-XRT than XMM-\emph{Newton} and \emph{Chandra} obtained in
$\sim300$\,ks of pointed observations of $z>5$ AGN. This observation is a
clear demonstration of concept for efficient X-ray studies of the high-$z$
IGM with large area, high-resolution X-ray detectors, and shows that
early-phase GRBs are the only backlighting bright enough for X-ray
absorption studies of the IGM at high redshift.

   \end{abstract}

   \maketitle

%
%--------INTRODUCTION---------
%
\section{Introduction\label{introduction}}
The promise of $\gamma$-ray bursts (GRBs) as cosmic lighthouses to
rival quasars is being fulfilled in the areas of GRB-DLAs
\citep{2004A&A...419..927V,2005astro.ph.10368W,2005astro.ph..8237S,2005astro.ph..8270C},
as tracers of star-formation
\citep{2005MNRAS.362..245J,2003A&A...406L..63F,2002AJ....123.1111B}, and as early warning of SNe
\citep[e.g.\ SN\,2003lw,][]{2004A&A...419L..21T,2004ApJ...608L..93C,2004ApJ...609L...5M}.
Central to this promise is the belief that GRBs from early in the
universe can be detected \citep[$z\sim10$, e.g.][]{2003ApJ...591L..91M}. But
while the highest redshifts of AGN and galaxies increased, 
for 5 years the highest GRB redshift was $z=4.50$
\citep{2000A&A...364L..54A}. Now, a GRB at $z>6$ has finally been
detected: GRB\,050904 at $z=6.295\pm0.002$ (\citeauthor{2005astro.ph.12052K}
\citeyear{2005astro.ph.12052K}, see also
\citeauthor{2005astro.ph..9660H} \citeyear{2005astro.ph..9660H}, \citeauthor{2005astro.ph..9766T}
\citeyear{2005astro.ph..9766T}, and \citeauthor{2005astro.ph..9697P}
\citeyear{2005astro.ph..9697P}).
To date, X-ray observations of $z>5$ AGN with \emph{Chandra} and
XMM-\emph{Newton} have obtained bare detections
\citep{2002ApJ...571L..71S,2002ApJ...569L...5B,2002ApJ...570L...5M,2003AJ....125.2876V,2003ApJ...588..119B},
and from the most luminous, spectra with a few hundred counts using long exposures
\citep{2004ApJ...611L..13F,2005astro.ph..1521G,2005ApJ...630..729S}, allowing contraints to be placed on AGN
evolution up to the edge of reionisation. In this \emph{Letter} we
examine the X-ray spectra and lightcurve of GRB\,050904 from
\emph{Swift}-XRT.
Uncertainties quoted are at the  90\% confidence level unless otherwise stated.
A cosmology where $H_0=70$\,km\,s$^{-1}$\,Mpc$^{-1}$,
$\Omega_\Lambda = 0.7$ and $\Omega_{\rm m}=0.3$ is assumed throughout.

\section{Observations and data reduction\label{observations}}
GRB\,050904 triggered \emph{Swift}-BAT at 01:51:44\,UT. The BAT and XRT data
were obtained from the archive and reduced in a standard way using the most
recent calibration files. The BAT spectrum is well-fit with a single
power-law with photon index $\Gamma = 1.26\pm0.04$ and 15--150\,keV fluence
= $5.1\pm0.2\times10^{-6}$\,erg\,cm$^{-2}$, consistent with early results
\citep{2005GCN..3910....1C,2005GCN..3918....1P} that also suggested a
duration $T_{90} = 225\pm10$\,s. An upper limit to the peak energy of the
burst, $E_{\rm peak} > 130$\,keV was found by fitting a cut-off power-law
model to the spectrum and deriving the 3$\sigma$ limit on the cut-off
energy. The \emph{Swift}-XRT rapidly localised a bright source
\citep{2005GCN..3920....1M} and began observations in windowed timing (WT)
mode at $\sim170$\,s after the trigger and photon counting (PC) mode at
$\sim580$\,s.

\section{Results\label{results}}

\begin{figure*}
 \begin{center}
 \includegraphics[angle=-90,bb=-50 68 555 750,width=1.0\textwidth,clip=]{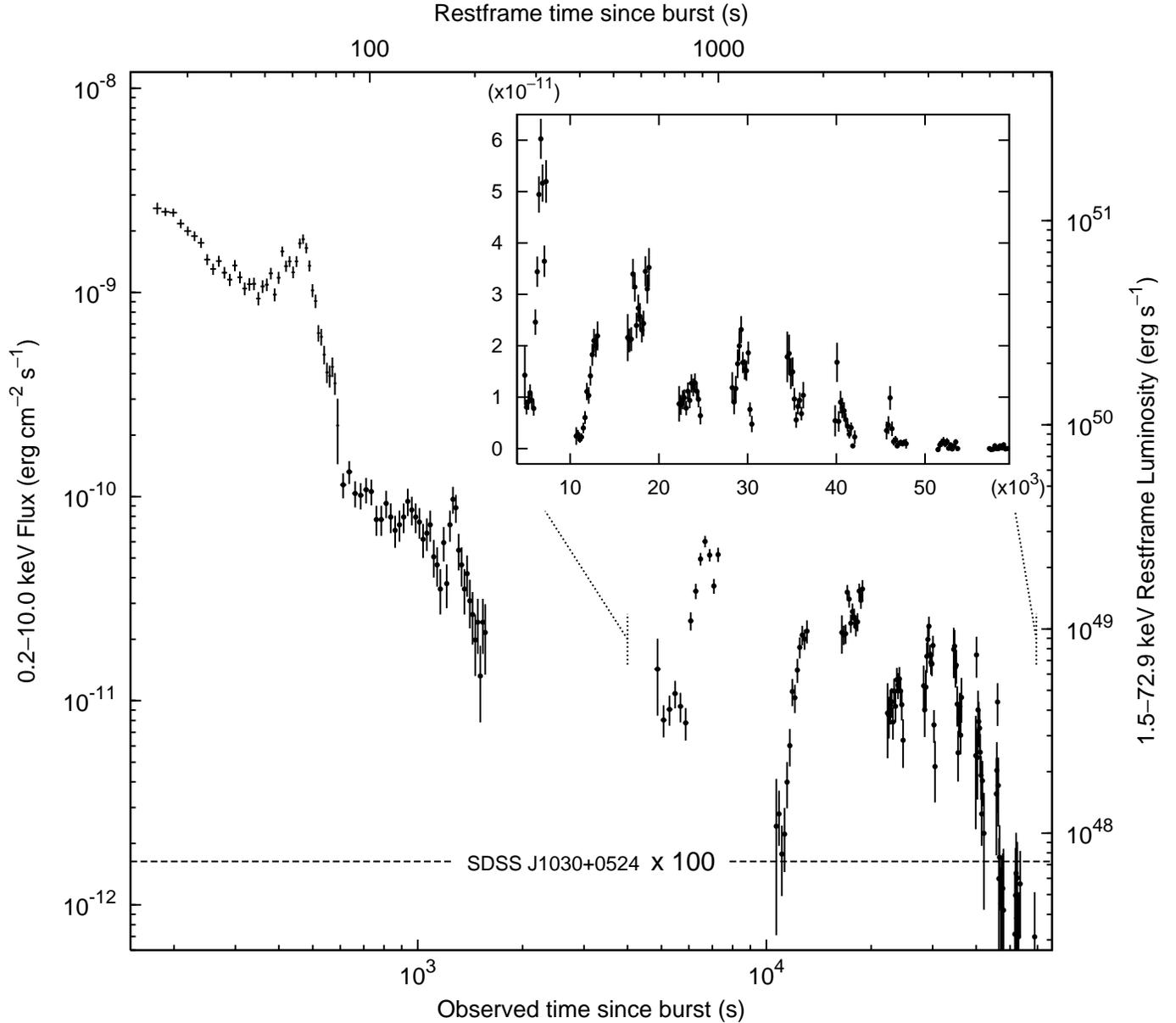}
 \caption{\emph{Swift}-XRT 0.2--10.0\,keV lightcurve of GRB\,050904
         ($\sim1.5$--72.9\,keV in the restframe). The equivalent isotropic
         luminosity at $z=6.29$ is plotted on the right axis. WT and PC mode
         data are indicated by crosses and dots respectively. The flux of
         one of the most luminous X-ray sources known, the AGN
         SDSS\,J1030+0524, is plotted for comparison. We have had to
         multiply its flux by 100 to get it on the plot.  SDSS\,J1030+0524
         was the most distant known X-ray source before GRB\,050904 and is,
         conveniently, at nearly the same redshift ($z=6.28$).
         \emph{Inset:} Linear blow-up of the data from
         $\sim10-70$\,ks to illustrate the variability of the source at late
         times. The very hard spectral index at early times
         ($\Gamma\sim1.2$) and the long BAT $T_{90}$ for this burst,
         indicate that most of the WT mode data is dominated by prompt
         emission. However, the continued large amplitude variability more
         than one and a half hours after the GRB trigger (in the restframe),
         and the still relatively hard spectrum ($\Gamma\sim1.9$), suggests
         that the XRT lightcurve is dominated by emission from the central
         engine during the first twelve hours of observations.}
 \label{fig:lightcurve}
\end{center}
\end{figure*}

\begin{figure}
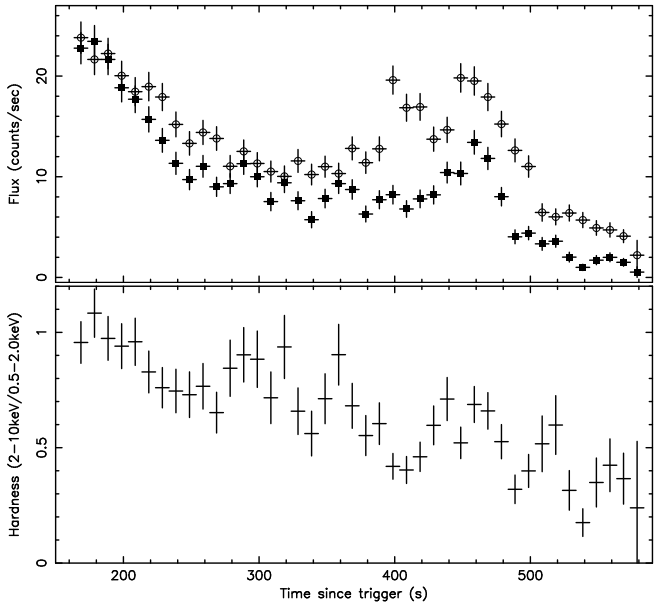

 \includegraphics[angle=-90,width=\columnwidth,bb=248 43 531 700,clip=]{f2a.eps}
 \includegraphics[angle=-90,width=\columnwidth,bb=249 43 571 700,clip=]{f2b.eps}
 \caption{\emph{Upper:} Soft (0.5--2.0\,keV, open circles) and hard (2--10\,keV, filled squares) early lightcurve of GRB\,050904.
          \emph{Lower:} Hardness ratio of the early lightcurve.  The hard to
          soft evolution, observed in most GRB prompt emission, is fairly
          monotonic outside the flares, where small deviations are
          discernible.}
 \label{fig:hardness}
\end{figure}

The XRT lightcurve (Fig.~\ref{fig:lightcurve}) 
fades by $>1000$ over the
first day. But the lightcurve does not decay as a power-law as in
many afterglows
\citep{2005astro.ph..8332N,2005astro.ph..7708D,2005astro.ph..7710G}.
Instead, the afterglow flares at $446\pm3$\,s, doubling the flux. This
flaring is similar to that observed in other GRBs at early times
\citep{2005Sci...309.1833B}, but the lightcurve does
not settle into a power-law decay, continuing to be dominated by large
variability (up to a factor of ten). The WT lightcurve is poorly fit by a
power-law plus a single Gaussian emission peak ($\chi^2$/dof = 195.7/78).
Allowing a second peak improves the fit (but is still poor, $\chi^2$/dof =
125.7/75), giving central times of $468\pm3$ and $431^{+5}_{-7}$\,s.
Dividing the data into hard (2--10\,keV) and soft bands (0.5--2.0\,keV) it
is clear that the later peak is harder, and the earlier peak softer
(Fig.~\ref{fig:hardness}). 
A two-peak fit to the soft band is acceptable
($\chi^2$/dof = 41.6/34) and gives different peak times than the fit to the
full band. There is considerable scatter around this model when fit to the
hard band data, giving an unacceptable $\chi^2$/dof (93.5/34), which
suggests greater variability in the hard band on timescales of $\sim10$\,s.

\begin{figure}
 \includegraphics[angle=-90,width=\columnwidth,clip=]{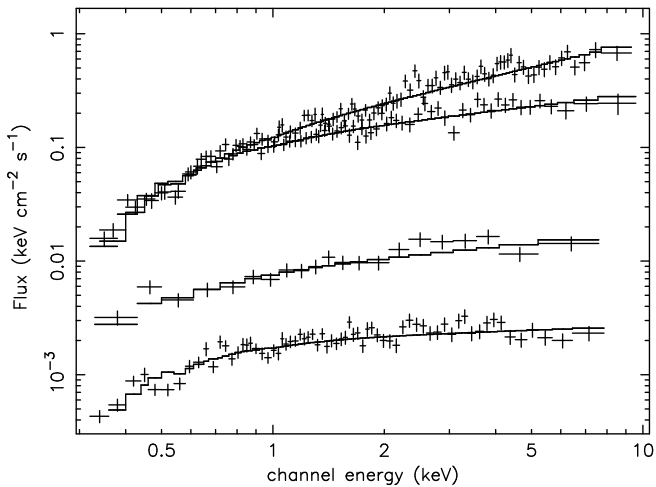}
 \caption{\emph{Swift}-XRT $E^2F(E)$ (equivalent to $\nu F_\nu$) spectra of GRB\,050904
          with the detector response removed. The spectra are fit with absorbed
          power-laws and show a clear hard to soft evolution, but the photon
          power-law indices are consistently $\Gamma<2$, suggestive of a
          decreasing peak energy that is above the bandpass ($\gtrsim70$\,keV
          in the restframe). The best fit parameters are listed in
          Table~\ref{tab:spectra}.}
 \label{fig:spectra}
\end{figure}

The spectra (Fig.~\ref{fig:spectra}) 
\begin{table}
\caption{Spectral evolution of GRB\,050904}
\label{tab:spectra}
\setlength{\tabcolsep}{6pt}
 \begin{center}
  \begin{tabular}{lcccccc}
   \hline\hline
   Mode	& Time since trigger (s)	&  $\Gamma$	& $N_{\rm H}$ at $z=6.29$\\
	& 				& 		& ($10^{22}$\,cm$^{-2}$)\\
   \hline
   WT	& 174--374			& $1.23\pm0.05$	& $3.3\pm1.5$	\\
   WT	& 374--594			& $1.62\pm0.06$	& $3.6\pm1.4$	\\
   PC	& 594--1569			& $1.68\pm0.08$	& $<1.6$	\\
   PC	& 9080--63480			& $1.88\pm0.04$	& $2.9\pm0.8$	\\
   \hline
  \end{tabular}
 \end{center}
\end{table}
can be fit by a hard power-law with
Galactic absorption
\citep[$4.9\times10^{20}$\,cm$^{-2}$,][]{1990ARA&A..28..215D}. The spectrum
softens appreciably during the observation, reaching $\Gamma\sim1.9$ in the
10--50\,ks after the GRB (Table~\ref{tab:spectra}). There is no evidence for
discrete emission or absorption features. \ion{Fe}{26}
(6.97\,keV) and \ion{Ni}{28} (8.10\,keV) at $z=6.29$ have respective restframe
equivalent widths $<43$ and $<44$\,eV in the WT spectra and $<27$ and
$<137$\,eV in the PC spectra. There is some evidence of absorption above the
Galactic level: the best fit gives
$N_{\rm H}=8.3\pm0.8\times10^{20}$\,cm$^{-2}$. This excess ($N_{\rm
H}=3.4\times10^{20}$\,cm$^{-2}$) is statistically required (significant at a
level $>5\sigma$ using the $f$-test). Typical variations in the hydrogen
column density at scales $\lesssim1\deg$ at high Galactic latitudes
are too small to explain this excess
\citep*{1990ARA&A..28..215D,1994ApJS...95..413E}. Without discrete features,
the redshift of the absorption is essentially unconstrained. Because
of the high redshift of the GRB, to observe even a modest column at $z=0$
requires a high column at $z=6.29$; in this case the best-fit excess column
density at $z=6.29$ is $2.8\pm0.8\times10^{22}$\,cm$^{-2}$. Such a high
column could not be considered entirely unexpected---a column density nearly
as high as this has been detected before in a GRB
\citep[e.g.][]{2005astro.ph.10368W}. Nonetheless it is intriguing at such an
early time in the star-formation history of the universe, especially since
the absorption is dominated primarily by oxygen and other $\alpha$-chain
elements. However, it should be noted that the combination of the
uncertainties in the Galactic column density and the current calibration
uncertainty of the XRT response at low energies must render one cautious
about the detection of excess absorption in this case.

\section{Discussion\label{discussion}}

The BAT-detected emission overlaps the start of XRT observations and
has a power-law photon index close to that observed in WT mode
($\Gamma=1.3$). It is likely that we are observing part of the prompt
emission with the XRT at these times
given the similarity with the BAT
spectrum, the rapid decay, the flaring, and a spectrum that softens
considerably over the first few hundred seconds in the restframe. This may
not be surprising considering the restframe energy band extends to nearly
$73$\,keV.
The fact that we are observing higher restframe energies in this GRB does not
seem to contribute much to the remarkable variability of the lightcurve, since
the soft band (0.5--2.0\,keV) has similar overall variability
(Fig.~\ref{fig:hardness}).  The amplitude of these variations seems to indicate
continued energy injection from the central engine at least for the first few
hundred seconds. Interestingly, the large variability continues as
late as 45\,ks (Fig.~\ref{fig:lightcurve}), and the spectrum remains hard
($\Gamma<2.0$), suggesting that significant energy output from the central
engine is likely to be continuing at these times, corresponding to
$\sim6000$\,s in the restframe. While continued energy injection at observed
times of up to a few hours has been indicated since the launch of \emph{Swift}
\citep{2005Sci...309.1833B,2005astro.ph..8332N}, energy injection from the
remnant at times of more than half a day was proposed to
explain the late-appearing X-ray line emission in GRB\,030227 \citep{2003ApJ...595L..29W,2000ApJ...545L..73R}.
The maximum heights of the later variations in GRB\,050904 also seem to decay
exponentially, indicating that if accretion onto the remnant is responsible for
these variations, that the accretion rate is decaying in the same way.

A power spectral density analysis of the lightcurve shows no significant
periodicity independent of the period of the data gaps in the range
$10^{-3}-10^{-4}$\,Hz. The large flaring amplitude and lack of a periodic
signal is reminiscent of typical prompt phase emission from GRBs. However,
the total duration of the flaring ($\gtrsim45$\,ks) and the individual rise
times (a few thousand seconds) are much longer \citep{2002A&A...385..377Q}.
The overall decay envelope observed here is not typical of prompt emission
either, although there are a few cases where such an overall decay is seen
(BATSE triggers 678, 2891, 2993, 2994, 7766) and it has been speculated that
these continuous decays of the prompt emission result from spin-down of a
black hole by magnetic field torques \citep{2002A&A...393L..15M}.

\subsection{Is GRB\,050904 Different?}

Assuming an upper limit to the redshift of GRB formation of $z=20$
\citep{2004NewA....9..353B}, the likely maximum age of the GRB progenitor is
$\lesssim650$\,Myr, consistent with a massive star progenitor
\citep{1998Natur.395..670G,2003Natur.423..847H,2003ApJ...591L..17S,2005astro.ph..8175W}.
At this early time in the universe, the question arises whether GRB\,050904
could have a different progenitor than GRBs at lower redshift; for instance,
a star formed in pristine gas may be one of the massive population\,III
stars.

Assuming the relation
between total energy ($E_\gamma$) and $E_{\rm peak}$
\citep{2004ApJ...616..331G}, the high restframe $E_{\rm peak}$ ($>940$\,keV)
implies a very high $E_\gamma$ \citep[$>2\times10^{51}$\,erg, consistent
with a possible jet-break in the near-infrared,][]{2005astro.ph..9766T}.
This high $E_\gamma$ and the large isotropic equivalent energy suggests that
GRB\,050904 was intrinsically highly energetic. The persistence of the flaring in the X-ray lightcurve,
is also different from typical GRB X-ray afterglows after a few hours
\citep{2005astro.ph..7710G,2005astro.ph..7708D}. Both the high intrinsic
energy output and the large amplitude, long duration flaring are notable
differences between GRB\,050904 and typical GRBs, and might hint at an
unusual progenitor. On the other hand, the X-ray flux of the afterglow at 10
hours, $\sim10^{-11}$\,erg\,cm$^{-2}$\,s$^{-1}$, implies a k-corrected
equivalent isotropic luminosity of $5\times10^{46}$\,erg\,s$^{-1}$, well within
the typical range \citep{2003ApJ...590..379B}. Although if the beaming
correction is relatively small, as suggested by the high value of $E_{\rm
peak}$, the energy inferred for the X-ray afterglow would also be large.

\subsection{High-$z$ Warm IGM Studies with GRBs}
Access to the edge of the reionisation epoch using GRBs has begun with the
observation of GRB\,050904 at $z=6.295$. Optical studies of the intervening
matter at early times have used quasars
\citep[e.g.][]{2001AJ....122.2850B,2001ApJ...560L...5D,2005ApJ...628..575W},
but may be affected by the quasar's significant influence on its
surroundings. GRBs are therefore potent tools in this study at optical
wavelengths. With X-rays, the warm intergalactic medium (IGM) can be probed.
Such work has also just begun to bear fruit with very bright,
nearby sources (e.g.\ \citeauthor{2005ApJ...629..700N}
\citeyear{2005ApJ...629..700N}, see also, \citeauthor{2002ApJ...572L.127F}
\citeyear{2002ApJ...572L.127F} and \citeauthor{2003ApJ...582...82M}
\citeyear{2003ApJ...582...82M}). This is because millions of X-ray photons
are required to make these absorption line measurements reliably
\citep{2000ApJ...539..532F,1998ApJ...509...56H}. The blazar Mkn\,421
($z=0.03$) has a bright, intrinsically featureless continuum which provides
an easily-modelled spectrum against which to detect intervening absorption
features. Long exposures ($\sim250$\,ks) with the gratings on \emph{Chandra}
provided $\sim7.5\times10^6$ photons from this source, mostly when the
blazar was in extremely bright flaring states. This allowed
\citet{2005ApJ...629..700N} to detect absorption from ionised
C, N, O, and Ne from IGM filaments at $z=0.011$ and $z=0.027$. The
spectra of GRBs, in prompt or afterglow emission, are usually dominated by a
featureless power-law \citep[although
see][]{2003ApJ...595L..29W,2003ApJ...597.1010B,2000Sci...290..955P,2002Natur.416..512R,2003ApJ...591L..91M},
as observed in this case, which makes them ideal for studies of intervening
matter in a way analogous to blazars
\citep{2000ApJ...544L...7F,2005astro.ph..4594K}.

The rapid response of \emph{Swift} to GRB\,050904 yielded high
signal-to-noise ratio X-ray spectra in spite of the relatively modest
aperture of the XRT. This contrasts favourably with observations with
\emph{Chandra} and XMM-\emph{Newton} of AGN at redshifts $z>5$ that have so
far yielded many fewer counts, even in aggregate, in a total exposure time of
300\,ks
\citep[e.g.][]{2001AJ....121..591B,2002ApJ...571L..71S,2002ApJ...569L...5B,2002ApJ...570L...5M,2003AJ....125.2876V,2003ApJ...588..119B,2004AJ....128.1483S,2004ApJ...611L..13F},
\emph{excluding} the deep field observations and in spite of the far larger
collecting areas of both instruments.

The $>10^{-9}$\,erg\,cm$^{-2}$\,s$^{-1}$ X-ray continuum detected in the
first minutes after GRB\,050904, demonstrates the power of GRBs to probe the
universe in X-rays to the highest redshifts. Follow-up observations of GRBs
with XMM-\emph{Newton}, and \emph{Chandra}, have shown that in practice the
typical fluxes for observations made more than $\sim6$ hours after the burst
\citep{2005astro.ph..7710G} are too low to detect the ionised IGM
\citep[c.f.][]{2000ApJ...544L...7F}. For instance, GRB\,020813, with one of
the highest average fluxes, provided about 5000 counts in the \emph{Chandra}
gratings over 100\,ks \citep{2003ApJ...597.1010B}; out of more than thirty
observations over the past five years, the average observed fluxes are
$\lesssim2\times10^{-12}$\,erg\,cm$^{-2}$\,s$^{-1}$. It is now clear that
observations of GRB afterglows with instruments not possessing a very rapid
response cannot provide grating spectra with anywhere near 100\,000 counts,
as had been speculated \citep{2000ApJ...544L...7F}. It is also now clear
that a good detection of the IGM requires a flux high enough to provide in
excess of $10^6$ counts at moderate spectral resolution. It would be
feasible to obtain enough photons in 50--100\,ks with the Narrow Field
Instruments on the proposed \emph{XEUS} mission, if it began observing up to
about 6 hours after the burst; with the brighter bursts this might also be
possible with \emph{Constellation-X}. But this is clearly not the most
efficient way to study the IGM with GRBs. It was suggested as an
alternative, that a high resolution instrument with small effective area
could make rapid observations of GRB afterglows in their early phases
\citep{2000ApJ...544L...7F}. To exploit the huge fluence provided by the high state and flares in the
first few minutes after the GRB, a small area detector could routinely
provide 10\,000 counts, but this is insufficient for IGM studies
\citep{2005ApJ...629..700N}. A very rapid response, similar to
\emph{Swift}'s, to a GRB like GRB\,050904 with a large area detector with
good spectral resolution and fast readout times (e.g.\
\emph{Constellation-X} or \emph{XEUS}) would reliably yield several to tens
of millions of photons in an exposure of only a few minutes. Such short
observations would allow studies of the high-redshift universe along many
different sightlines; observations that would each require months of
\emph{effective} exposure time observing a high-$z$ AGN. Such rapid
observations are demanding, but this is a technique now demonstrated in
practice by \emph{Swift}, and the short observations would be highly
efficient as well as providing superb spectra. A sample of such observations
could allow us to fix the fraction of baryonic dark matter, determine the
metallicity and density evolution of the IGM, and put strong constraints on
structure formation at high redshifts.

\begin{acknowledgements}
The Dark Cosmology Centre is funded by the DNRF. We acknowledge benefits
from collaboration within the EU FP5 Research Training Network, `Gamma-Ray
Bursts: An Enigma and a Tool'. We are indebted to S.~Larsson for
the power spectral density analysis.
\end{acknowledgements}

\section*{}
After submission of this \emph{Letter}, a paper by
\citeauthor{2005astro.ph..9737C}, appeared on the arXiv preprints servers
(astro-ph/0509737) based on the analysis of the XRT and BAT data from
GRB\,050904. Their findings are similar to those reported here.

\end{document}